\documentclass[aps,twocolumn,prl,amsmath,amssymb,superscriptaddress]{revtex4-1}

\usepackage{graphicx}
\usepackage{bm}
\usepackage{hyperref}
\usepackage{IEEEtrantools}

\usepackage{lineno}
\usepackage{natbib}
\usepackage{float}

\restylefloat{table}

\def\cm{cm$^{-1}$}
\def\kHgCl{k-Hg-Cl}

\begin{document}

\title{Melting of charge order in the low-temperature state of an electronic ferroelectric.} 

\author{Nora M. Hassan}
\affiliation{Institute for Quantum Matter and Department of Physics and Astronomy, Johns Hopkins University, Baltimore, MD 21218, USA}

\author{Komalavalli Thirunavukkuarasu}
\affiliation{Department of Physics, Florida A \& M University, Tallahassee, FL, 32307, USA}

\author{Zhengguang Lu}
\affiliation{National High Magnetic Field Laboratory, Tallahassee, Florida 32310, USA}
\affiliation{Florida State University, Tallahassee, Florida 32306, USA}

\author{Dmitry Smirnov}
\affiliation{National High Magnetic Field Laboratory, Tallahassee, Florida 32310, USA}

\author{Elena I. Zhilyaeva}
\affiliation{Institute of Problems of Chemical Physics, Chernogolovka, Russia}

\author{Svetlana Torunova}
\affiliation{Institute of Problems of Chemical Physics, Chernogolovka, Russia}

\author{Rimma N. Lyubovskaya}
\affiliation{Institute of Problems of Chemical Physics, Chernogolovka, Russia}

\author{Natalia Drichko}\email{Corresponding author: drichko@jhu.edu}
\affiliation{Institute for Quantum Matter and Department of Physics and Astronomy, Johns Hopkins University, Baltimore, MD 21218, USA}

\date{\today}
\maketitle



{\bf Strong electronic interactions can drive a system into a state with a symmetry breaking. Lattice frustration or competing interactions tend to prevent a symmetry breaking, leading to quantum disordered phases.  In spin systems frustration can produce a  spin liquid state.
Frustration of a charge degree of freedom also can result in various exotic states, however, experimental data on these effects is scarce. In this work we demonstrate how a charge ordered ferroelectric  looses the order on cooling to low temperatures using an example of a Mott insulator on a weakly anisotropic triangular lattice $\kappa$-(BEDT-TTF)$_2$Hg(SCN)$_2$Cl. Typically,   a  low temperature ordered state is  a ground state of a system, and the demonstrated re-entrant behavior is unique.
Raman scattering spectroscopy  finds that this material enters an insulating ferroelectric ``dipole solid'' state at $T=30~K$, but below $T=15~K$ the order melts, while preserving the insulating energy gap. The resulting phase diagram is relevant to  other quantum paraelectric materials.}

Frustration of a charge degree of freedom can result in charge glass~\cite{Sato2017,Sasaki2017} or a quantum paraelectric state, where electric dipoles fluctuate down to the lowest temperatures~\cite{Yao2018,Shen2016,Hotta2010}. Such quantum dipole liquid was observed experimentally in a band insulator on a triangular lattice~\cite{Shen2016} and in a Mott insulator~\cite{Hassan2018}. In a band insulator fluctuations of polarization are
 predicted to lead to a multiferroic effect~\cite{Dunnett2019}.  In a Mott insulator charge-spin coupling is predicted to result in a spin liquid state~\cite{Hotta2010,Naka2016}.
An experimental realization of a system  where electrical dipoles form on lattice sites of  a Mott insulator at this point is limited to molecular-based systems~\cite{Yao2018,Hotta2010,Hassan2018}. However, exotic multiferroicity~\cite{Dunnett2019,Naka2016magnetoelectric}  which can result from an interplay of a quantum paraelectric and a spin liquid states is of interest to a broad community working on materials with  strong electron-electron interactions.   Also, notable is an analogy of a charge degree of freedom on the orbital of molecular dimer (BEDT-TTF)$_2$ to the orbital degree of freedom and orbital liquid in atomic crystals, and as a way to produce novel spin liquid states~\cite{Naka2016}.

Organic Mott insulators where electronic ferroelectricity and quantum dipole liquid are observed are layered charge-transfer crystals based on BEDT-TTF~\footnote{bis(ethyleneditio)tetrathiafulvalene} molecule. Layers responsible for the interesting physical properties of these materials are formed by dimers (BEDT-TTF)$_2^{+1}$. They alternate with layers which serve as charge reservoirs and define the exact structural parameters of the BEDT-TTF layers. A compound discussed in this work $\kappa$-(BEDT-TTF)$_2$Hg(SCN)$_2$Cl (\kHgCl) shows a structure where   (BEDT-TTF)$_2^{+1}$ sites form a slightly anisotropic triangular lattice within the layer. In this compound   electronic ferroelectricity is observed in a charge ordered state below 30~K~\cite{Drichko2014,Gati2018}. In this work we experimentally detect a gradual melting of this charge order as the material is cooled down below 15~K.

Typically, if a system undergoes a phase transition into a broken symmetry state, such as a ferroelectric state, this state is a ground state of the system.  A loss of an order on lowering temperature, for example,  in a  re-entrant transition, is very rare to observe experimentally.  Such a re-entrant transition was predicted theoretically for some cases of strong electronic interactions~\cite{Merino2003,Murray2013,Neilson2013,Watanabe2017}.
In quasi-two-dimensional (quasi-2D) organic Mott metals a weak re-entrant behavior was observed  close to insulating charge order phase for compounds with a quarter-filled conductance band~\cite{Drichko2006}.

\section{Results}

We use Raman spectroscopy as a main tool for this study. This method allows us to access molecular vibrations to characterize charge distribution on the BEDT-TTF-formed lattice,  phonons, and electronic excitations, providing full information about the state of the material and a symmetry of the order.


\begin{figure*}
  \includegraphics[width=18cm]{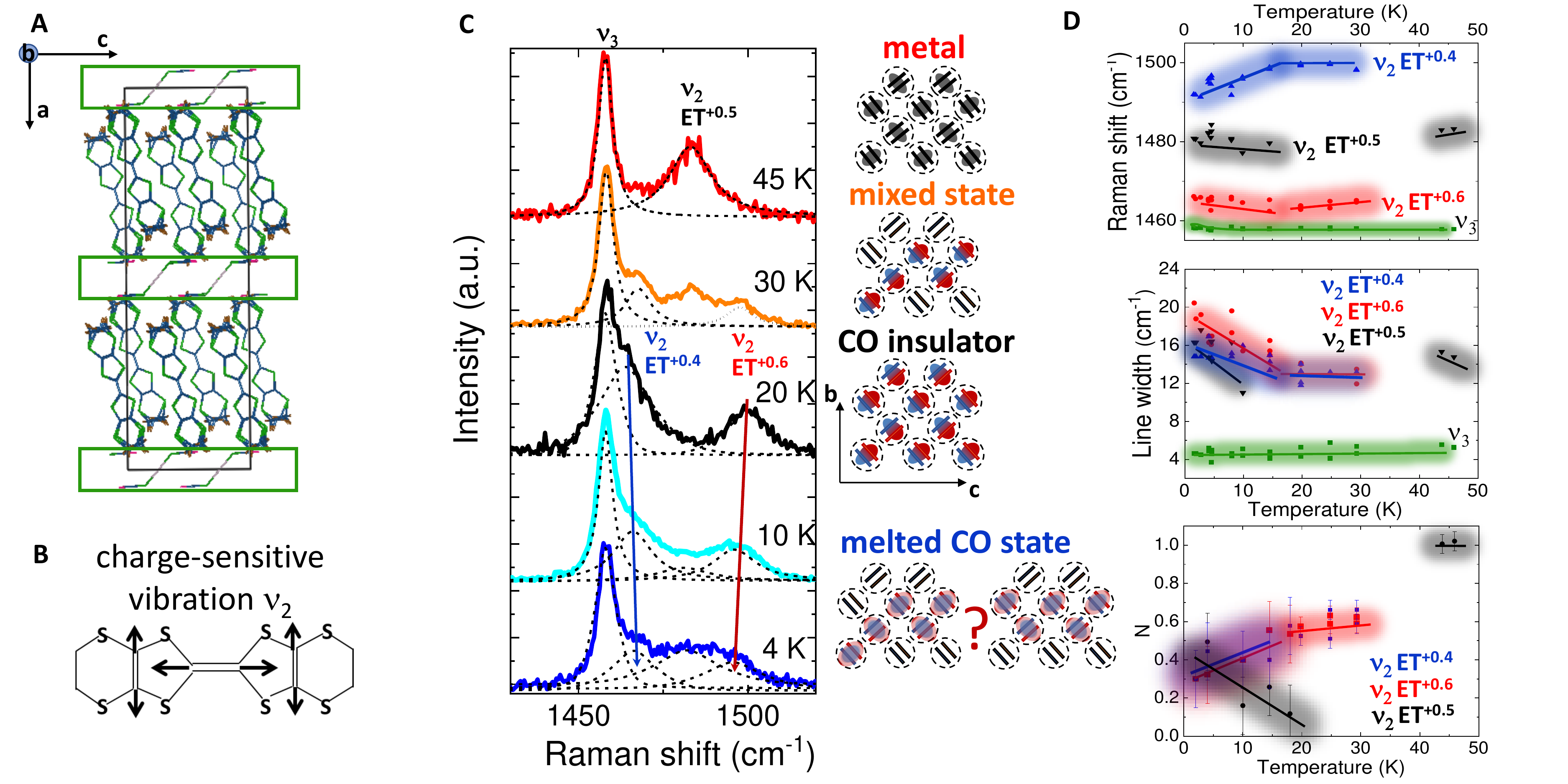}\\
  \caption{{\bf (A)} Layered crystal structure of quasi-two-dimensional (quasi-2D) molecular-based Mott insulators. {\bf (B)} Eigenvectors of charge-sensitive molecular vibration $\nu_2$. {\bf (C)} Raman spectra of $\kappa$-ET-Cl in the spectral range of charge-sensitive vibration $\nu_2$ at different temperatures. Schemes at the right present the charge distribution on the lattice, black is for BEDT-TTF$^{+0.5}$, red for BEDT-TTF$^{+0.6}$, and blue is for BEDT-TTF$^{+0.4}$; ET = BEDT-TTF on the plots.  From the top to the bottom: In the metallic state at 45~K only $\nu_2$(BEDT-TTF$^{+0.5}$) is observed; At the temperature of charge order transition T$_{CO}$=30~K mix of  metallic domains with  $\nu_2$(BEDT-TTF$^{+0.5}$)  and charge ordered domains with $\nu_2$(BEDT-TTF$^{+0.6}$)/ $\nu_2$(BEDT-TTF$^{+0.4}$) is observed; In the charge ordered state at T=20~K  $\nu_2$(BEDT-TTF$^{+0.6}$) and  $\nu_2$(BEDT-TTF$^{+0.4}$) are observed; At T=10~K charge order starts to melt,  broadened $\nu_2$(BEDT-TTF$^{+0.6}$)\ $\nu_2$(BEDT-TTF$^{+0.4}$) and weak $\nu_2$(BEDT-TTF$^{+0.5}$) are observed, a fraction of BEDT-TTF$^{+0.5}$ is about 0.2 of the whole system;  At T=4~K  broadened $\nu_2$(BEDT-TTF$^{+0.6}$)\ $\nu_2$(BEDT-TTF$^{+0.4}$) and  $\nu_2$(BEDT-TTF$^{+0.5}$) are observed, about 0.3 of the system has each charge state.    {\bf (D)} Temperature dependence of the frequencies (upper panel) and line width (middle panel) of $\nu_2$ and $\nu_3$.  Note the splitting of $\nu_2$ band below 30~K, and then broadening and change of frequencies which reduces the difference between $\nu_2(A)$ and $\nu_2(B)$. The lower panel shows a temperature dependence of the fraction of the system {\bf N} associated with different charges on BEDT-TTF molecule.  }\label{Vibrational}
\end{figure*}


First we will discuss the charge ordered state which appears in \kHgCl\ below the metal-insulator transition at 30~K~\cite{Drichko2014}.  The charge distribution in the layer formed by (BEDT-TTF)$_2^{+1}$ molecular dimers is determined  by following the frequency of the Raman-active $\nu_2$ stretching vibration of the central C=C bond of the BEDT-TTF molecules, which has a known linear dependence on charge on the molecule~\cite{Hassan2018,Yakushi2015}. A single $\nu_2$ mode observed above the transition (Fig.~\ref{Vibrational}, 45~K) splits into two components found at 1475 and 1507~\cm\ (T=27~K) in the spectra of \kHgCl\ below   30~K. The latter corresponds to a charge ordered state where charge symmetry within each dimer lattice site is broken, with  (BEDT-TTF)$^{+0.6}$ and (BEDT-TTF)$^{+0.4}$ at each dimer (see Fig.~\ref{Vibrational}).


\begin{figure*}
  \includegraphics[width=16cm]{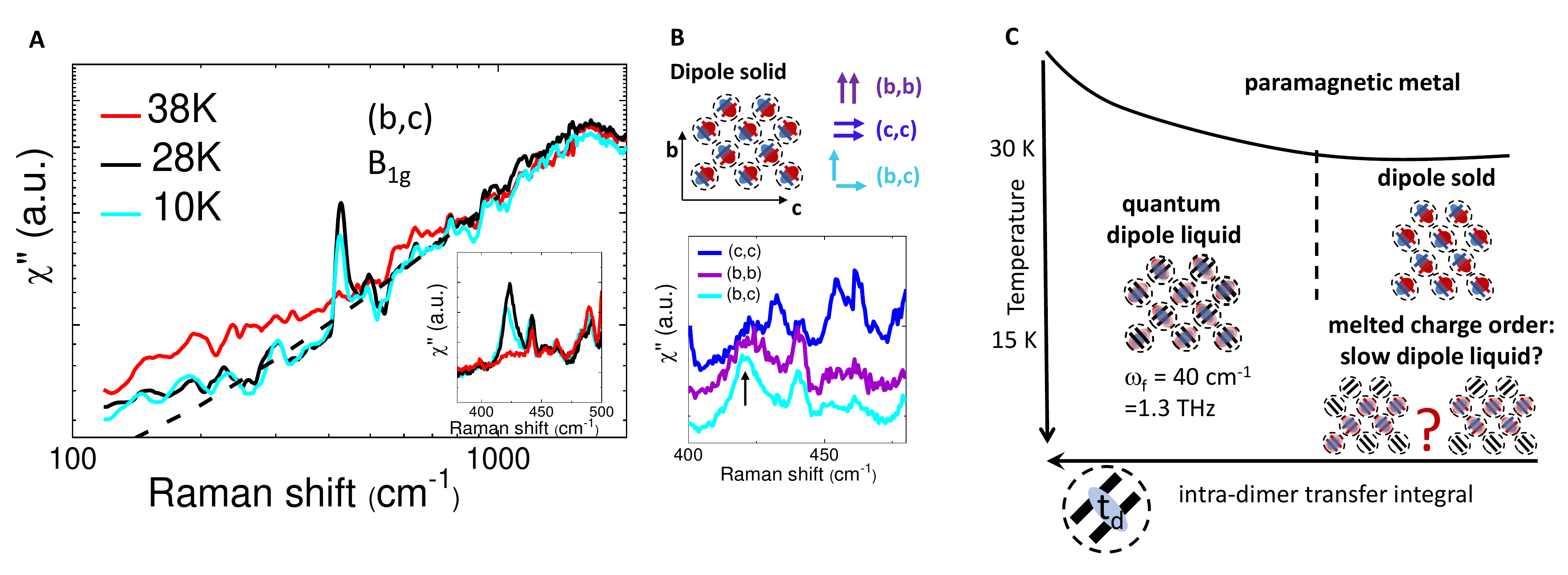}\\
  \caption{({\bf A}) Electronic scattering of \kHgCl\ in B$_{1g}$ scattering channel in metallic state (T=38~K), charge ordered state (T=28~K), and at T=10~K with approximately   0.16 of the charge order melted. Note the gap of 250~\cm\ which opens on the transition into the charge ordered state and preserved at T=10~K. The inset shows an ``amplitude mode'' at these temperatures. Note the decrease of intensity at changing the temperature from 28 to 10~K. ({\bf B}) Upper panel: Scheme of light polarization orientation to the lattice of \kHgCl\ with charge stripes along $c$ axis. Lower panel: polarization dependence of the feature at about 420~\cm, which appears in the spectra in the charge ordered phase and is an analogue to an amplitude mode in spectra of charge density wave. Note the anisotropy of the mode. ({\bf C}) Suggested phase diagram where overlap integral in a dimer t$_d$ controls a quantum phase transition between dipole (charge) order and non-charge-ordered Mott insulator phase. A phase transition on lowering temperature is of the first order (solid line), while a quantum phase transition controlled by t$_d$ is the second order transition (dashed line). }\label{Gap}
\end{figure*}


Vibrational features in the spectra of \kHgCl\ are superimposed on a continuum of electronic excitations, presented in  Fig.~\ref{Gap} after subtracting phonon contributions.  In  $(b,c)$ scattering channel, on the M-I phase transition at 30~K  we observe an opening of an electronic gap. It manifests itself in a suppression of the intensity of the continuum below approximately 850~\cm\ with spectral weight shifted to the region between 850 and 1200~\cm.  The size of the gap 2$\Delta$ can be estimated as a frequency where the slope of $\chi''(\omega)$ flattens due to the suppression of spectral weight (Fig.~\ref{Gap}A). This estimate yields 2$\Delta$ of about 250~\cm. An opening of an insulating gap due to electronic correlations in Raman response is observed in strongly correlated materials, for example in SmB$_6$ ~\cite{Valentine2015,Devereaux07}. The $(b,c)$ scattering channel   corresponds to B$_{1g}$ scattering channel in the spectra of  cuprate superconductors, where charge order effects are well studied~\cite{Caprara2015}.

The 1D properties of the charge  stripes  formed in the charge ordered state~\cite{Drichko2014,Gati2018}  are evidenced by a  broad asymmetric feature that appears at about 400~\cm\ in $(b,c)$,  and $(b,b)$ scattering channels,   while its intensity in $(c,c)$ is negligible (see Fig.~\ref{Gap}).
The symmetry and  position of this feature is reminiscent of an amplitude mode of a charge density wave, which would be found below an electronic gap and reveals the anisotropy of the order~\cite{Eiter2013}. While the exact details of the origin of this feature are outside of the scope of this paper, its position suggests that it originates from a coupling of electrons to a vibration of BEDT-TTF molecule.\footnote{$\nu_{10}(A_{1g})$ vibration is known to couple to localized electrons~\cite{Eldridge1995}, and with additional symmetry breaking would appear both in Raman and in infrared spectra~\cite{Drichko2014} in the crystallographic direction along which charge is alternating in the structure.}

An important question about the ordered phase in \kHgCl\ is  the strength of coupling of the electronic ferroelectricity to the lattice. In a strong coupling regime a formation of a spin singlet between the charge-rich sites is suggested~\cite{Dayal2011}. In the absence of a coupling to the lattice the charge-rich stripes can be considered as 1D antiferromagnetic (AF) chains~\cite{Watanabe2017} and have a potential to show 1D spin liquid properties. No lattice symmetry breaking was found by XRD measurements by now~\cite{Drichko2014}.
Raman spectroscopy study also does not observe any major changes   for lattice phonons which lie below 100~\cm, confirming  the absence of major structural changes. Some small changes in the molecular vibrations of BEDT-TTF on the phase transition reflect the charge redistribution (for details see SI).

Typically, a broken symmetry state would be a ground state of a system. The situation is different for \kHgCl\, where we observe melting of the charge order below  approximately 15~K. The first evidence comes from vibrational spectroscopy: The components of the charge sensitive band $\nu_2$ start to gradually broaden,   move closer together in frequencies, and  loose spectral weight to the increasing  $\nu_2$ (BEDT-TTF$^{+0.5}$). Basing on the temperature dependence of the intensity of  $\nu_2$ band for different charges on BEDT-TTF molecule (see Methods for details) we can estimate a dependence of fractions $N$ of  differently charged  BEDT-TTF molecules  on temperature, presented in Fig.~\ref{Vibrational}B, lower panel.

At T=10~K the estimated distribution is:\\ $N(+0.4e):N(+0.6e):N(+0.5e)$  = 0.4$\pm$0.15 : 0.4$\pm$0.15 : 0.16$\pm$0.15. The spectral weight of the 400~\cm\ in-gap mode decreases proportionally to the part of the system that lost charge order: The spectral weight $I(10~K)= 0.82 I(20~K)$ is in agreement with the vibrational spectra estimate of a charge order loss in about 0.16$\pm$0.15 of the total number of molecular dimers.   While loosing spectral weight, the feature at 400~\cm\  preserves the anisotropy relevant to the charge stripes along $c$ axis. The Mott electronic gap at 250~\cm\ does not  change below the charge order transition at 30~K (Fig.~\ref{Gap}A).
No major changes on vibrational features apart from $\nu_2$ are observed, confirming again that the coupling of the charge order to the lattice is weak.

At the lowest measured temperature of 2~K it is difficult to distinguish single components of  $\nu_2$ mode, which shows intensity spread from about 1450 to 1510~\cm. If we identify the components of the $\nu_2$ in a way similar to the 10~K data analysis,  we can estimate that the charge distribution is:\\ $N(+0.4e):N(+0.6e):N(+0.5e)$ = 0.3$\pm$0.15 : 0.3$\pm$0.15 : 0.3$\pm$0.15.

The shape of the $\nu_2$ line in the low temperature phase of \kHgCl\ is reproduced on different cooing cycles, and for cooling rates varied between  0.1K/min to 40 K/min in measurements with the laser probe sizes of 50*100~$\mu$m and 2~$\mu$m.  Thus, if the mixture is formed by domains, they are much smaller than 2~$\mu$m.

The the width of $\nu_2$ line components increases gradually of on cooling into the melting regime. This is in contrast to the behavior of a mixed state of metallic and charge ordered domains  observed at the temperature of the charge order phase transition T=30~K (see Fig.~\ref{Vibrational}B). The spectra of the mixed state show a superposition of the charge-homogeneous and charge ordered response without additional line broadening. Such a formation of local strain-controlled domains on a phase transition into charge ordered insulating phase was observed in $\alpha$-(BEDT-TTF)$_2$I$_3$~\cite{Pustogow2018}. Typically, domains formed at a phase transition in BEDT-TTF-based materials vary in size from about 1~$\mu$m ~\cite{Pustogow2018} to 100~$\mu$m~\cite{Sasaki2004}.

\section{Discussion}

As our data show,  an interplay between electronic correlations and frustration of the lattice leads to exotic effects related to the metal-insulating transition and insulating state. In comparison to a classic example of a charge-ordered insulating transition in $\alpha$-(BEDT-TTF)$_2$I$_3$ at  T$_{MI}$=135~K~\cite{Ivek2011} with electronic gap of 600~\cm, \kHgCl\ shows a larger gap of   2$\Delta$=8kT. The charge order in \kHgCl\  is weakly coupled to the lattice, which explains an absence of  a spin singlet state~\cite{Yamashita_kHgC}.

The main result of our study is that the charge order   observed at temperatures below T=30~K is  not the ground state of the system.   On cooling below 15~K charge order gradually melts with an appearance of BEDT-TTF$^{+0.5}$, while the fraction of the system which still holds charge order shows broadened charge-sensitive $\nu_2$ lines moving closer in frequency on cooling. This gradual melting of charge order   does not have a detectable effect on heat capacity~\cite{Hassan2018}. Also, an absence of a dependence on the cooling rate does not favour an interpretation in terms of charge glass, in agreement with an absence of glassy behavior in dielectric measurements~\cite{Gati2018}. The melting of  charge order occurs without a formation of macroscopic domains of different phases and can result in a charge-fluctuating or microscopically disordered charge state.

 Basing  on vibrational spectroscopy only it is not possible to distinguish a decrease of charge disproportionation accompanied by some additional charge disorder from an offset of slow charge fluctuations. However, a continuous change of the line positions and width on cooling is in agreement with charge fluctuations scenario.  A frequency of charge fluctuations
or, in other words, dipole fluctuations in this case is about 4~\cm\ at 2~K, according to the line width analysis~\cite{Hassan2018}. This is an order of magnitude slower than dipole fluctuations observed in the quantum  dipole liquid $\kappa$-(BEDT-TTF)$_2$Hg(SCN)$_2$Br.
The fraction of the system  which preserves the charge order also preserves 1D anisotropy, as measured by anisotropic Raman response at 420~\cm.
We estimate that at the lowest measured temperature of 2~K each charge state presents approximately 1/3 of the system.

While our results is the first report  of  the low-temperature melting of the dipole order in \kHgCl, some published experimental data for this material reveal this effect. ESR data~\cite{Yasin2012,Gati2018} show a decrease of $\chi_M$  at the charge order transition, with a recovery of the metallic state value of $\chi_M$  below T=20~K. This behavior can be associated with the melting of charge order or a related loss of magnetic short range correlations between charge-rich sites. Resistivity of \kHgCl~\cite{Drichko2014,Gati2018} increases by a few orders of magnitude on the metal-insulator transition at 30~K, but the  temperature dependence flattens below  20~K. Both results suggest that below approximately 20~K the system changes the behaviour and becomes ``less insulating'',  while the metallic behaviour is not fully recovered. Interestingly, for these lower-frequency probes the upturn of the behavior occurs at slightly higher temperatures than our observations, which would agree with interpretation of the low-temperature state in terms of fluctuations.

There are a number of models which can explain our observation on the melting of the charge order in  the low-temperature phase of \kHgCl:

A re-entrant transition is predicted for a system with electronic ferroelecticity of molecular dimers in Ref.~\cite{Naka2010}. It is suggested that antiferromagnetic exchange on lowering the temperature would compete with the interactions which lead to ferroelectric order, and weaken them, leading to the loss of the ferroelectric order.

A number of models for a frustrated charge system discuss a competition between a stripes charge order similar to that in \kHgCl\ and  different variants of a three-fold charge order.  This phase competition at low temperatures can result in a melting of a ferroelectric charge order state without an impact of magnetic interactions.  Ref.~\cite{Yoshida2014}  discusses  a competition between change order stripes and an "order by disorder" phase with so-called "good defects" containing ordered pairs produced by thermal fluctuations. Another possibility is a competition between ferroelectric stripes and anti-ferroelectric three-fold stripes ground state discussed in Refs.~\cite{Mori2016,Watanabe2017}. Interestingly, a competition between ferroelectric and antiferroelectic state is a reason for quantum paraelectric behavior in SrTiO$_3$~\cite{Chandra2017}.

Our experimental data present a limited agreement with these models, without suggesting a clear picture. At 2~K we  observe a  charge distribution which corresponds to a three-fold charge order, but it is not clear if this charge distribution is the final state of the system or it would  change on cooling. The 1D anisotropy is preserved in the melted state for the fraction of the system which holds charge separation. This result can agree both with a competition of different stripe orders and  with microscopic domains containing stripes, which would appear on a re-entrant transition. Three-fold charge order is a metallic state, and the stripe order of this kind is antiferroelectric and thus would change the dielectric constant behavior.  Further experimental studies such as detailed dielectric measurements  and resistivity below 30~K are necessary for the identification of the low temperature state.

The models above work in the assumption that the FE state is static. Since the dielectric measurements are done at about 1000 GHz, and Raman and optical measurements are done at even higher frequencies, it cannot be excluded that very slow fluctuations, with frequency below 100 GHz are present even between 30 and 20 K.   This would suggest that the MI transition and the gap at 250~\cm\ is  related to purely Mott transition.

Our results not only demonstrate the unique effect of melting of an ordered state at low temperatures in \kHgCl\, but also suggest a  phase diagram drafted in Fig.~\ref{Gap}C. A transition into insulating state on temperature lowering is of the first order both for \kHgCl~\cite{Gati2018} and for the closely related quantum dipole liquid $\kappa$-BEDT-TTF$_2$Hg(SCN)$_2$Br~\cite{Hassan2018,Hemmida2018}. The fluctuations of the charge order in $\kappa$-BEDT-TTF$_2$Hg(SCN)$_2$Br and dipole ordered phase or possibly much slower fluctuations in \kHgCl\ suggest a quantum critical point and a second order phase transition into a charge ordered state. According to both theory~\cite{Hotta2010,Naka2010,Sato2017_Ishihara} and DFT~\cite{Gati2018,Valenti_private} the tuning parameter of the ground state is  an overlap integral $t_d$ in a (BEDT-TTF)$_2$ dimer~\cite{Hotta2010,Hassan2018}.  Interestingly, a bandwidth controlled phase diagram suggested by calculations in Ref.~\cite{Sato2017_Ishihara} also has a 1st order phase transition between metallic and antiferromagnetic insulating phase, while a transition between metal and antiferromagnetic phases, and electric dipole ordered phase is 2nd order. A similar phase diagram with 1st order phase transition with temperature, and 2nd with another parameter, contain  triple quantum critical point is also suggested for ferroelectrics such as SrTiO$_3$ without spin degree of freedom~\cite{Chandra2017}.

\section{Acknowledgements}

The authors are grateful to T. Clay, T. Mori, C. Hotta, H. Seo, M. Naka, T. Ivek, M. Yamashita for fruitful discussions. Work in JHU was supported as part of the Institute for Quantum Matter, an Energy Frontier Research Center funded by the U.S. Department of Energy, Office of Science, Office of Basic Energy Sciences under Award Number DE-SC0019331. The work in Chernogolovka was supported by FASO Russia, state registration number 0089-2019-0011. KT acknowledges funding via ONR HBCU/MI program. A portion of this work was performed at the National High Magnetic Field Laboratory, which is supported by the National Science Foundation Cooperative Agreement No. DMR-1644779 and the State of Florida. ND and NH acknowledge the support of Institute for Complex Adaptive Matter (ICAM) and The Gordon and Betty Moore Foundation.

\bibliography{OrganicConductors2018full}

\section{Methods}

{\bf Synthesis}
Single crystals of $\kappa$-(BEDT-TTF)$_2$Hg(SCN)$_2$Cl (\kHgCl)  were prepared by electrochemical oxidation of the BEDT-TTF solution in 1,1,2-trichloroethane (TCE) at a temperature of 40$^{\circ}$ C and a constant current of 0.5~$\mu$A. A solution of Hg(SCN)$_2$, [Me$_4$N]SCN·KCl, and dibenzo-18-crown-6 in 1:0.7:1 molar ratio in ethanol/TCE was used as supporting electrolyte for the \kHgCl\ preparation.  The composition of the crystals was verified by electron probe microanalysis and X-ray diffraction.

{\bf Raman scattering} was measured in three different setups.

(i) In the frequency range from 100 to 2000~\cm and temperatures down to 10~K spectra were measured in the pseudo-Brewster angle geometry using T64000 triple monochromator spectrometer equipped with the liquid N$_2$ cooled CCD detector. In single monochrmonator configuration with the edge filter option was used. Spectral resolution was 2~\cm. Line of Ar$^+$-Kr$^+$ Coherent laser at 514.nm was used for excitation. Laser power was kept at 2~mW for the laser probe size of approximately 50 by 100~$\mu$m. This ensured that laser heating of the sample was kept below 2~K, as was proved by observing the temperature of ordering transition in \kHgCl. Measurements at temperatures down to 10 K were performed using Janis ST500 cold finger cryostat. Cooling rates used were between 0.1 and 40~K/min. The samples were glued on the cold finger of the cryostat using GE varnish. The experiments were performed on at least 6 samples to ensure reproducibility of the results.

Non-polarized micro-Raman measurements in the range of charge-sensitive vibrations were performed in backscattering geometry using T64000 triple monochromator spectrometer equipped with the Olympus microscope. Probe size was 2 $\mu$m in diameter.

The crystals were oriented using polarization-dependent  Raman scattering measurements.
For the measurements, electrical vector of   excitation $e_L$ and scattered $e_S$ light were polarized along $b$ and $c$ axes. Our notations of polarizations refer to the structure and symmetry of the BEDT-TTF layer, to make an easy comparison to the calculations which refer to D$_{4h}$~\cite{Devereaux07} without loosing the information about the symmetry of the real crystal. Thus A$_{1g}$ symmetry corresponds to the  were measurement in $(b,b)$ and $(c,c)$ geometries,  and B$_{1g}$ corresponds to $(b,c)$ and $(c,b)$ geometries ($xy$). 
All spectra were corrected by the Bose-Einstein thermal factor.

(ii) Measurements down to 1.9~K in the spectral range between 1200 and 1700~\cm\ with a spectral resolution of 1.74~cm$^{-1}$ were conducted in NHMFL at 0~T. The Raman spectra were measured in a backscattering geometry using a 532nm laser excitation. The laser light was injected into an single mode optical fiber, guiding the excitation to the sample stage inserted into a helium-flow variable temperature cryostat. The excitation light was focused by an aspheric objective lens(NA=0.67) to a spot size of about 3.5 $\mu$m in diameter. The excitation power delivered to the sample was about 100$\mu$W or less to minimize the sample heating. The scattered light collected by the same lens was directed into a 100 $\mu$m multimode collection fiber, and then guided to a spectrometer equipped with a liquid-nitrogen-cooled CCD camera. The spectra were acquired in the spectral region from 1200  cm$^{-1}$ to 1700 cm$^{-1}$  with a spectral resolution of 1.74 cm$^{-1}$.

{\bf Data analysis}

We determine charge on BEDT-TTF molecules in the structure by following the frequency of the central C=C molecular bond  vibration ($\nu_2$) (Fig.~1B,C). This frequency changes by $\sim$ -140~\cm\ when  the charge  of the molecule changes from (BEDT-TTF)$^0$ to
(BEDT-TTF)$^{+1}$, as demonstrated in Refs.~\cite{Dressel04,Yamamoto2005}. This is a result of a lengthening of the central C=C bond of the molecule when more charge occupy the highest occupied molecular orbital (HOMO).

\begin{figure}
  \includegraphics[width=8cm]{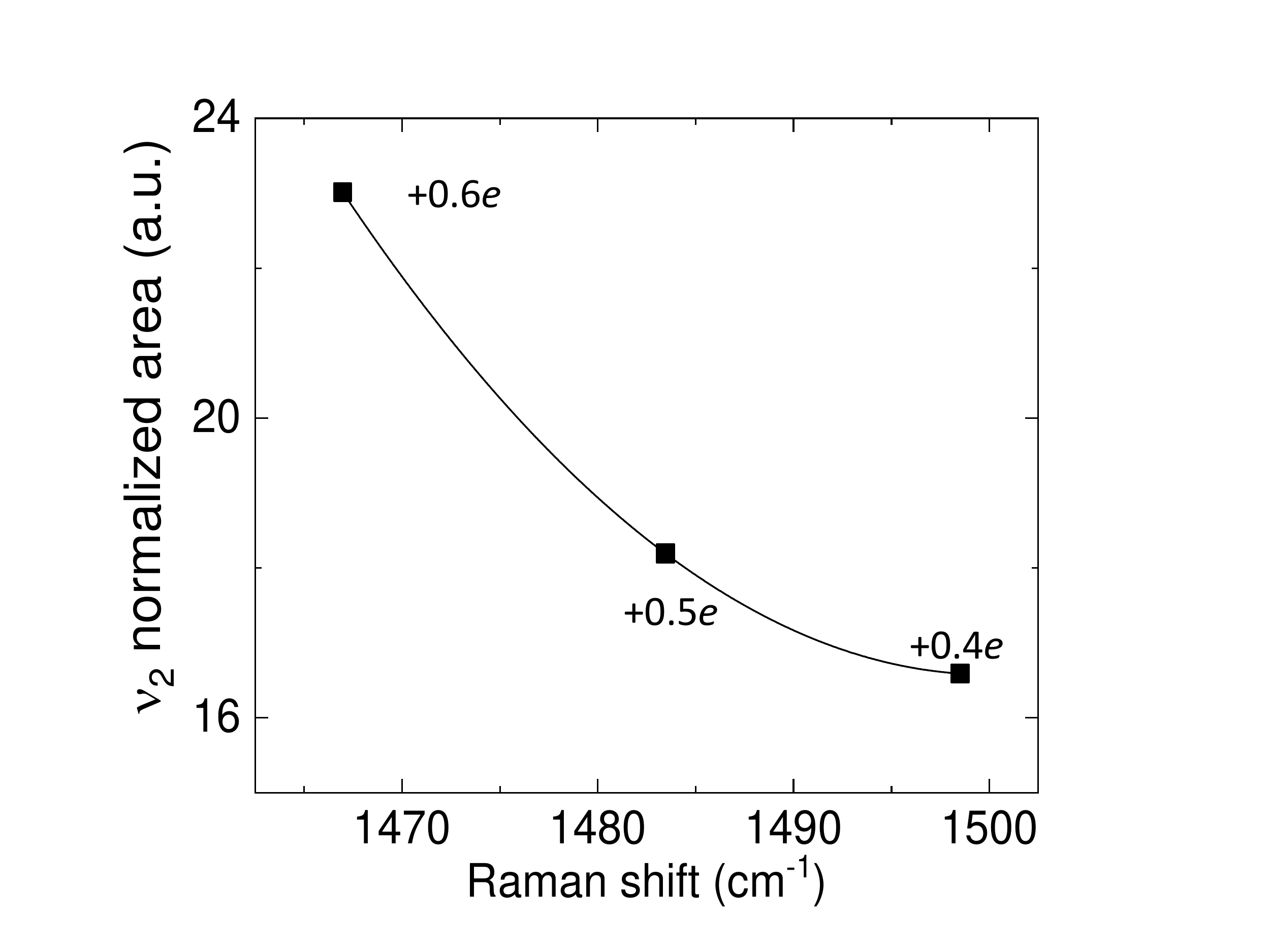}\\
  \caption{Dependence of normalized intensity of a charge sensitive mode $\nu_2$ of BEDT-TTF molecule on Raman shift. The latter is defined by the charge on the molecule. Points represent the known values of BEDT-TTF average charge. }\label{Methods}
\end{figure}

To compute the fractions of the system $N$ which carries certain charge at temperatures below 15~K, where charge order starts to melt (see  Fig~\ref{Vibrational}D lower panel) we performed a comparative analyses of Raman intensities. While frequency of the $\nu_2$ charge sensitive vibration linearly depends on the charge on BEDT-TTF molecule~\cite{Yamamoto2005}, the Raman intensity shows a non-linear dependence, which we obtained from the experiment. The known points of this dependence is T=45~K in the metallic state, where all molecules have average +0.5$e$ charge, and T=22~K, where half of the system carriers charge +0.6$e$, and half is +0.4$e$. Spectral weight of $\nu_2$  vs Raman shift plot is presented in Fig.~\ref{Methods}. Used in the plot spectral weight values are received from the fits of the vibrational bands with Lorentzian shape and are normalized to the spectral weight of $\nu_3$ C=C non-charge sensitive band to correct for the errors of intensity measurements at different temperatures, which can in principle occur due to small misalignment of a system at changing temperature.  We used the empirical dependency (solid line) to estimate the fraction $N$ of BEDT-TTF molecules carrying a certain charge at temperatures below 15~K.

\section{Supplemental Information}

\subsection{Vibrational spectra of the mixed state at the charge order transition temperature  vs low temperature charge order melted state.}

We follow the charge state of \kHgCl\ on cooling through the charge order phase transition by following Raman active charge sensitive $\nu_2$ vibration~\cite{Yakushi2015}. A single $\nu_2$ mode observed above the transition (Fig.~\ref{Vibrational}, 45~K) splits into two components found at 1475 and 1507~\cm\ (T=27~K) in the spectra of \kHgCl\ below   30~K suggests charge ordered state with charge  disproportionation $\delta n$= 0.2$e$. At the temperature of the phase transition T=30~K we observe a mix of charge order and homogeneous charge response (see Fig.~\ref{Vibrational}). The widths of the vibrational bands in the mixed state are similar to that in the metallic state just above the transition and the ordered state below 30~K. This suggest at the mixed state consists of macroscopic domains, and does not introduce additional disorder on the microscopic level. Since the ratio of charge-homogeneous to charge ordered fraction of the system is similar on different cooling cycle, we expect domains to be much smaller than the 2~$\mu$m radius of the laser probe.

Domains can be a signature of the 1st order phase transition which was suggested for this compound~\cite{Gati2018}. In addition, since thermal conductivity of the materials decreases on a transition into the insulating state, the domains can be associated with small inhomogeneities of cooling of the crystal on the cold finger of the cryostat.

Importantly, a comparison of the vibrational spectra in the mixed state with the spectra in the low temperature charge melted state reveal important differences. The broadening on the bands at low temperatures suggest charge fluctuations or disorder or microscopic  charge disorder of the fraction of the system which preserved separation of changes on the lattice at low temperatures.

\begin{figure}
  \includegraphics[width=9cm]{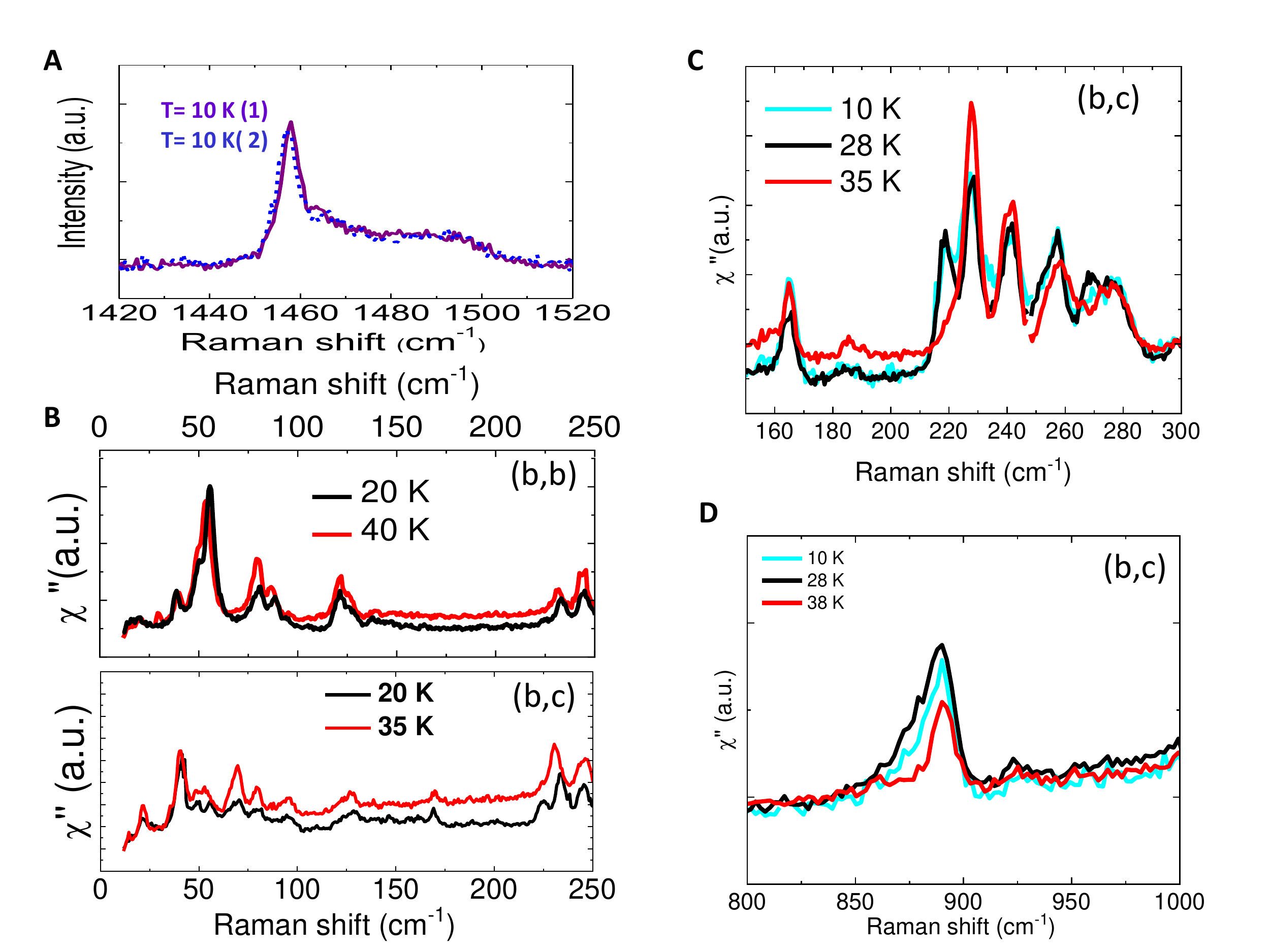}\\
  \caption{Vibrational spectrum of $\kappa$-(BEDT-TTF)$_2$Hg(SCN)$_2$Cl (A) Micro-Raman (probe diameter 2 $\mu$m) spectra in the spectral region of $\nu_2$ charge sensitive vibration measured at the same position on the crystal on two successive cooling cycles; (B) Spectra in the spectral range of phonons (12 to 250~\cm) above the charge order transition at 30~K (T=40 and 35 K) and in the charge ordered phase at 20~K. Note the absence of major changes on the phase transition, which corresponds to the absence of the lattice changes. (C-D) Spectra in the region of   BEDT-TTF molecular vibrations at temperatures above and below the ordering transition and in the melted phase at 10~K}\label{SI}
\end{figure}

\subsection{Vibrational changes on the CO phase transition}

While we do not observe and  notable changes in the lattice phonons on the charge ordering phase transition at 30~K in \kHgCl. However, in addition to $\nu_2$ some other lines of molecular vibrations of BEDT-TTF split on the phase transition. These vibrations listed below and shown in Fig~\ref{SI}B are associated with the deformations of the fragments of the aromatic rings of BEDT-TTF. Molecular $\pi$-orbitals of BEDT-TTF have a large input of orbitals of the aromatic rings, and a change of geometry of these rings on the change of the charge could be expected.

We observe a doubling of a vibrational lines at about 228~\cm, which is assigned to a B$_{3g}$ deformation of S-C-S angle between the two aromatic rings  and an angle between central C=C and C-S of the first ring, and the one at about 290~\cm, which is related to and A$_{1g}$ stretching of the bonds of the outer aromatic ring~(Fig.~\ref{SI}B, upper panel)~\cite{Kozlov1987}.

We also observe an enhanced intensity of the vibrational line at about 900~\cm\ in the charge ordered state. This band is related to a B$_{3g}$ stretch of a C=C  bonds of the inner rings, and was claimed to couple to electronic transitions in BEDT-TTF materials~\cite{Eldridge1996}


\end{document}